# Experimental Evidence for the Attraction of Matter by Electromagnetic Waves

Hans Lidgren

Le Mirabeau, 2 ave des Citronniers, MC 98000, Monaco

Rickard Lundin

Swedish Institute of Space Physics, Box 812, 981 28 Kiruna, Sweden

rickard.lundin@irf.se

ABSTRACT

We present measurement results demonstrating that a conducting lead sphere exposed to electromagnetic (e/m) waves in the infrared (IR) regime, is attracted by e/m waves. The result may seem surprising and against conventional wisdom that electromagnetic wave forcing should lead to a repulsive force. Nonetheless, all our experiments show that the attractive force can be determined quantitatively, and that they are reproducible. Our experiment setup is a Cavendish torsion-balance experiment with lead spheres, one of the spheres intermittently irradiated by IR light. Because the Cavendish experiment is well known, simple, and readily available, the results can be easily verified or falsified. However, to avoid Bernoulli and other external forcing effects, the entire experimental setup should be placed in a vacuum chamber. In our case the experiments were performed at $\approx 4 \cdot 10^{-7}$ mbar. One of the 20 g lead spheres was intermittently irradiated by infrared radiation from a lamp covered by an aluminium foil. Two independent experiments (V1 and V2) are described. Besides showing that wave energy and momentum transfer leads to attraction, we also describe some experimental requirements and constraints. The lamp was powered on during 10 or 12 s, the power changing between 8, 16, and 26 W. All measurements, including those affected by lamp out-gassing, shows that the attracting force on the lead sphere increases with increasing irradiative energy. From the V2 experiment, preceded by lamp "baking" to eliminate repulsive out-gassing forces, the irradiative energy 8.7 Ws on the sphere resulted in a total force $2.9 \pm 0.5 \cdot 10^{-9}$ (N). From the V2 experiment we also derive a power law relation between incident radiation energy (W) and the attractive force, corresponding to $F = 2.8 \cdot 10^{-10} W^{1.1}$, with $R^2 = 0.95$.

*Subject headings:* Electromagnetic wave forcing; mass-attraction; Cavendish experiment

# 1. Introduction

A rigid body irradiated by electromagnetic waves is subject to forcing. On a macroscopic scale, electromagnetic wave forcing can result in heating, energy and momentum transfer. Electromagnetic (e/m) waves illuminating a surface may be reflected, become absorbed, and/or lead to secondary wave emission. The surface properties of a body, defined by their absorption-, emission-, and albedo coefficients, are generally frequency-dependent.

A Poynting flux describes the propagation of e/m waves in a certain direction. However, inherent in such a description is that because e/m waves move in a certain direction, wave-forcing results in momentum transfer in the same direction. This may be referred to as Maxwell's radiation pressure, i.e. e/m energy and momentum transfer results in a repulsive force on the body, here denoted the Repulsive Radiation Pressure-force (RRP-force).

Attempts to quantify and experimentally verify the RRP-force dates back to the late 19th and early 20th century. Maxwell presented the theory of radiation pressure in 1871. Thirty years later an experiment by Lebedeev (1901) came close by 20% of the theoretical value, the reflected radiation from a thin foil exerting a larger repulsive force than the absorbed radiation. A follow-up foil experiment (Nicholz and Hull, 1903) in low vacuum provided even better agreement with the radiation pressure theory. Contemporary research on radiation forcing utilizes primarily coherent (laser) light, consider effects on the atomic scale. This has lead to applications in low-temperature physics (Wineland and Itano, 1987, Cohen-Tannodji et al., 1990) and nanotechnology (Fedosov et al., 2007). However, macroscopic effects of radiation pressure on solid bodies are mainly considered as an RRP-force, like on satellites (Hughes, 1977). In view of the results presented here, the theory of solar light forcing of satellites should be reconsidered. The lack of studies on solid bodies may be due to the fact that the feeble radiation force on heavy targets is notoriously difficult to measure. Cavendish torsion balance experiments to determine the gravitational constant must be designed to avoid "anomalous" external influences, but under less controlled environments they should have been subject to the external forcing described here. As a result the torsion balance technique to test gravity in laboratories have been very carefully designed to eliminate as much as possible external and internal induced perturbations, in particular those caused by radiation/temperature effects. Results from advanced gravitational tests are now available, making it possible to determine violations from gravity down to separation distances of $10^{-8}$ m (Grundach, 2005).

Considering the "unconventional" way the forcing described in this report works, it is only natural for physicists to associate it with environmental effects such as Bernouilles airflow. A proper forcing experiment therefore requires not only vacuum, but also a controlled electromagnetic environment. The apparatus and environment used in this experiment does not match the requirements required for gravity experiments. However, first of all our experiments are performed in vacuum. Furthermore, the electromagnetic force applied in our experiment is higher compared to the feeble gravitation force between two lead balls in a gravitation experiment. In fact, our apparatus and environmental constraints are quite adequate for testing the impact of electromagnetic radiation forcing. Finally, the experiment set up is simple enough for any trained physicist to confirm or falsify our results in most physics laboratory with access to a vacuum chamber.

Neither classical e/m field theory nor quantum field theory implies that e/m-radiation leads to an attractive force of neutral bodies. However, the physics of magnetized plasma provides an alternative avenue in this respect, because plasma wave-induced ponderomotive forces (e.g. Allan et al., 1990, Li and Temerin, 1993) may lead to attractive forces under special circumstances (Lundin and Guglielmi, 2006).

## 2. Experimental Results

The experimental setup, a Cavendish torsion balance experiment, has been modified to enable determination of the radiation force caused by an encapsulated lamp placed in front of one of the lead spheres. To avoid potential air-streaming (Bernoulli) effects, the experiment is placed in vacuum at $\approx 4 \cdot 10^{-7}$ mbar. The radiation force from the encapsulated lamp was then derived from the pendulum offset. By encapsulating the light source by aluminium foil, the radiation becomes absorbed by the foil, the "lamp" now radiating isotropic in the infrared regime.

The Cavendish Unit from Leybold Didactic Gmbh, is a torsion pendulum consisting of two (20g ± 0.5g) lead spheres, placed on the ends (c/c 100 mm) of a light weight aluminium "boom", and suspended in the centre by a 26 cm long torsion band made from bronze. The pendulum period is 616 ± 3 s. This yields a torsion coefficient of $9.81 \cdot 10^{5}$ (Nm/radian). The "boom" is mounted inside a vacuum chamber of glass. The pendulum is equipped with a concave mirror. An infrared light pointer, 75±1 cm from the mirror, along with an IR position detector, indicates the oscillation of the pendulum. The detector enables the values to be automatically recorded and plotted in real-time by a computer. A plotted displacement of 1 mm is equal to a true motion of 0.067 mm. The centre of a 35 W halogen lamp, covered by aluminium foil, is placed 15±2 mm in front of one of the lead spheres. The power-on period of the lamp varies from 10 to 12 seconds. A diagrammatic representation of the radiation experiment using a Cavendish torsion pendulum is shown in figure 1.

The vacuum system, from Edwards High Vacuum International, is a turbomolecular pump backed by a rotary vane vacuum pump. Included in the system is also a nitrogen-cooling unit. The system provides a vacuum of ≈$3.4 \cdot 10^{-7}$ mbar. The equivalent mass of the energies used in the experiments, maximum 35 W for 20 seconds, is less than $10^{-14}$ kg and hence neglected.

Using only one radiation source in the experiments will result in asymmetric forces. Initially, only the lead sphere in front of the radiation source will be affected. It will initially also have a small effect on the axis of rotation of the balance and increase the oscillation period. These effects have been considered in the results. Electromagnetic forcing will also depend on the changing distance between the lamp and the lead sphere. The radiation decay versus time of the lamp has consequences for the experiment, as will be discussed later.

Forces between charges are strong and may affect the experiment results. Charges between the lead sphere and the lamp affect the oscillation amplitude. Charges also often change the equilibrium position of the balance. It was possible to detect when charges destructively interfered with the results of the experiments. The Cavendish torsion balance has been connected to the ground during all experiments. It has also been discharged when charges have been detected, or at least once every day when experiments were performed.

In the vacuum experiment, the lamp was placed inside the vacuum chamber. If the lamp were switched on after more than 6-12 hours of inactivity, there would be a flow of gas molecules towards the pendulum spheres, pushing the closest pendulum sphere away from the lamp. To decrease the effects of these repulsive forces, the lamp was "baked" at high temperatures before performing an experiment.

As a lamp in vacuum only looses energy by radiation, experiments in vacuum are affected by radiation much longer than under normal air pressure. The decaying radiation from a lamp therefore affects the experiment during a much longer time period, typically several tens of minutes, but also (experiment V2) for at least an hour. Differences in decay time between the two vacuum experiments presented here are mainly related with the lamp+foil. Experiments using an "unbaked" lamp+foil (V1) lead to gas releases and faster lamp decay. The two experiments are therefore not only independent of each other, but they have different radiation decay curves, and the V1 experiment demonstrates the effect a repulsive gas flow has on the experiment.

The oscillation of the torsion pendulum is essentially un-damped in vacuum due to extremely small frictional losses. Together with the need of "baking" the lamp before carrying out the experiments, the pendulum was rarely at rest at equilibrium when placed in vacuum. The pendulum is either accelerated or decelerated by the lamp radiation. Depending of the pendulum phase when the lamp was switched on, towards or away, the pendulum amplitude will be either damped or enhanced. Furthermore, because of the lamp radiation decay, forcing may extend over several pendulum cycles, leading to alternating acceleration and deceleration forcing. Consequently, the net increase/decrease of the pendulum amplitude depends on the radiation decay curve, the quicker the decay time, the larger the instantaneous forcing, and the higher the increase or decrease in pendulum amplitude. Finally, the net irradiation also depends on the pendulum amplitude, i.e. the distance between the radiation source and the lead ball. Because of the physical size of the radiation source, the target illumination does not go with $1/r^2$, where r is the distance between source and target, but rather with $1/r$ for small r. Acceleration towards the radiation source, and alternating forcing by the radiation decay is readily discernable in the experimental results (Figs 2 and 3).

In experiment V1, see Fig. 2, the influence of the radiation of gas molecules from the aluminium foil of the lamp during an experiment was tested. No "baking" was done during the last 12 hours before this experiment was performed. In this experiment, the lamp was illuminated 3 times. First the lamp power was on for 10 seconds at 5V, i.e. ≈9W (≈90 Ws). After approximately one hour, the lamp power was on again for 10 seconds but now at 10V, i.e. ≈26W (≈260 Ws). After approximately another two and a half hours, the lamp power was on for the third time during 10 seconds but now only at 8V, i.e. ≈18W (180 Ws). Notice from the pendulum amplitude response for three irradiation periods that radiation prevails for at least one pendulum period. Moreover, the more energy stored in the lamp foil, the longer the time for significant lamp forcing.

The lamp decay time may in fact be determined from the decaying pendulum amplitude. For experiment (1) we get the following expressions for the lamp decay:

$I = I_o \cdot \exp(-0,0039/t)$, where $I_o$ may be derived from the total energy acquired ($P \cdot \Delta t$), i.e. by the electric power (P) during a time $\Delta T$. A normalized response of the lamp based on the above time constant is depicted by red curves in Fig. 2. Dashed lines give the power on time for the three cases.

The first period with lamp power-on, 10 seconds at 5V, i.e. ≈9W nominal power, started immediately after the pendulum sphere, *leaving* the lamp, passed equilibrium. This illumination reduced the initial amplitude-maximum by 3mm. Notice that radiation of gas molecules from the aluminium foil of the lamp apparently reduced the amplitude shift of the attractive radiation force of the lamp during the first illumination period. When the pendulum sphere oscillates to the lamp-side, the amplitude is about the same as in the latter maxima. In this case, the radiation of gas molecules from the aluminium foil of the lamp apparently compensated for the attractive radiation force of the lamp.

The second power-on period, 10 seconds at 10V, i.e. ≈26W nominal power, also started immediately after the pendulum sphere, *leaving* the lamp, passed equilibrium. This reduced the initial amplitude-maximum by 9mm. When the pendulum sphere oscillates back to the other side, the initial amplitude of the pendulum sphere increased by 2mm. In this case, the attractive radiation force of the lamp was stronger than the counteracting radiation of gas pressure from the lamp. When the pendulum sphere turns to the maximum amplitude the next time, the decayed lamp radiation decelerates the pendulum sphere, decreasing the amplitude by 1mm. After that no more forcing effects, was observable. Notice the deflection difference between the first and second irradiation period, the first period affected more by outgassing effects than the second period.

The third power-on period, 10 seconds at 8V, i.e. ≈18W nominal power, started immediately after the pendulum sphere, now *approaching* the lamp, passed equilibrium. In this case the pendulum amplitude increased substantially, reaching its highest maximum. Moving towards the next minimum, the lamp continuous to attract the pendulum sphere. Notice the pendulum asymmetry with respect to the equilibrium line, the minima being closer to equilibrium than the maxima. The subsequent maxima decrease reflects a combination of lamp decay and increasing distance to the lamp. Conversely, the minute difference between subsequent minima is a consequence of a greater distance to the lamp. The radiation of the lamp affects the system for almost 15 minutes with half of the initial increase of the amplitude remaining in the system. The reason for this is that the second illumination of the lamp at 10V, i.e. ≈26 W nominal power, served as bake-out of gas molecules at a higher power level, with less gas molecules remaining at the lower (≈18 W) power level.

In experiment V2, "baking" of the lamp was done at 12V, i.e. 35W. To minimize the radiation of gas molecules from the aluminium foil of the lamp, during the experiment, the lamp was illuminated at only 10V, i.e. ≈26W nominal effect, see figure 3. The power-on period of the lamp was 12 seconds, starting when the pendulum sphere passed equilibrium *approaching* the lamp. As we can see in Fig. 3, the pendulum sphere more than doubled its amplitude after the power-on period. However, when the pendulum sphere oscillates back again, the attractive force of the lamp, still hot and radiating, decelerates the pendulum sphere and restores almost exactly its initial amplitude. When the pendulum sphere turns at its maximum amplitude again, the still-radiating lamp accelerates the pendulum sphere, increasing its amplitude again. The radiation of the lamp affects the system for essentially an hour, with half of the initial increase of the amplitude remaining in the system. From the pendulum response in the V2 experiment we may now determine the decay of the radiation source versus time. From Fig. 4, illustrating the pendulum displacement versus time, we may determine the radiation decay of the lamp. We note that the pendulum displacement versus time fits very well ($R^2=0.91$) with an exponential decaying curve, characteristic of lamp-cooling. Notice that this curve only applies to experiment 2. For experiment 1 the exponential decay falls off more rapidly. From the lamp displacement curve we may now derive the radiative power on the sphere versus time, noting that the total energy (312 Ws) should corresponds to the integral of the radiation curve versus time: $I(t) = I_0 \exp(-at)$   $t \geq t_0$, where $I_0$ is the radiated power at $t=t_0$ and $a$ is the decay constant. Setting $t_0 = 0$ in the V2 experiment we obtain: a = 0,0011 (s$^{-1}$); $I_0 \approx 0.34$ (W). From of the pendulum displacements observed and using the torsion coefficient of the pendulum (9,81·10$^5$ Nm/radian), we can derive the radiative forcing in the V1 and V2 experiments. Considering first the displacements from experiment V2 (312 Ws), we derive the force versus time from the displacement values in Fig. 4. The average accumulated force $F_k$ derived for each time interval $\Delta t$, is plotted in Fig. 5. Least square fitting to an exponential function gives: $F_k=9.3·10^{-10}\exp(-0.00113·t_k)$ with $R^2=0.91$. The total force is achieved by accumulating the force values in Fig. 5. In a similar manner we can from the average force derive the force versus time, i.e. F(t)=3.5·10$^{-12}$ exp(-0.0012·t), with R$^2$=0,93.

Notice that individual measurement values in Fig. 5 represent the average forces accumulated over a time interval, the radiation force acting in an accelerating- (towards the lamp) or decelerating (away from the lamp) sense. The distance to the lamp is here assumed to be constant, which means we have not taken into account for the larger attraction near maximum compared to the weaker attraction near minimum (further away from the lamp). The influence of the distance to the lamp is modelled in the companion paper (Lundin and Lidgren 2010). The sum of the averaged forces in Fig. 5, and the integral force derived from the exponential function F(t), give the following results:

Sum of force measurements (fig. 5): $\sum_{k=1}^{10} F_k = (2.3 \pm 0.5) \cdot 10^{-9}$ (N)

Integral force (exponential curve): $\int_0^{4000} F(t)dt = (2.9 \pm 0.6) \cdot 10^{-9}$ (N)

Under the assumption that the lamp is a point source that radiates isotropic, the radiation intensity will decrease by 1/r$^2$, where r is the distance between the lamp and the sphere. In reality the lamp has a finite size, and the radiation will reach the sphere at various distances and angle of incidences. Taking into account for the above we derive a geometric factor G=0.028 that should be applied to the total radiative power and energy input to the sphere. This means that for the V2 experiment the total energy input to the sphere is 8.74 J. The corresponding energy input for the V1 experiment is 2.49, 8.74, and 5.98 J, respectively. Based on the total true pendulum deviations observed from V1: 0.58, 1.23, and 1.10 mm respectively we derive in a similar way as for the V2 experiment the following forces: 5.9·10$^{-10}$, 1.2·10$^{-9}$, and 1.1·10$^{-9}$ N, respectively. The four total force values (V1 and V2), together with the partial force values (Fig. 5) are plotted in Fig. 6. Circles in Fig. 6 mark V1 values. Notice that the V1 force corresponding to 5.98 J, falls well below the integral value for V2. This is consistent with the fact that the V1 experiment was done without lamp baking; the repulsive force from gas/material ejection from the lamp+aluminium foil opposing the lamp attraction force. Nevertheless, a clear trend from the four individual forcing experiments is obtained, i.e. the total force increases with total energy received by the sphere, the power law curve ($F=2.8 \cdot 10^{-10} \cdot W^{1.1}$) suggestive of a linear relation between radiative energy input and total force.

## 3. Conclusions

A series of Cavendish experiments, two of them discussed in more detail here, demonstrate quite convincingly that e/m radiation may exert an attractive force on matter. The experiment is surprisingly simple, a Cavendish torsion pendulum setup, the pendulum comprising two 20 g lead spheres, one of the spheres illuminated by infrared radiation. Unambiguous results are obtained when the Cavendish experiment is placed in vacuum. The vacuum in our experiment, ≈$3.4 \cdot 10^{-7}$ mbar, implies that airflow can be completely discarded. Only radiative forces can significantly influence the lead-spheres in vacuum. Without radiative forcing, the pendulum moves essentially unperturbed for many hours. The lack of convective (air) cooling in vacuum implies that a clean (baked) radiation source (the lamp) cools very slowly. For instance, the radiation from the source in experiment V2 decreased exponentially with a decay constant ≈830 s. Notice that the long forcing period implies attraction that works in two ways, with and against the torsion spring force. The net force after seven oscillations (Fig 3) is therefore the result of a series of consecutive acceleration and deceleration periods.

We estimated the total force on the pendulum based on the total power input (e.g. 12W during 12 s) and the subsequent decay of the radiation source (up to seven oscillation periods). A least square power law fit of the total intergrated-, and partial (within Δt) forces give the following equation describing the attractive force versus electromagnetic energy input:

$$F = 2.8 \cdot 10^{-10} \cdot W^{1.1} \qquad R^2 = 0.95$$

The V1 data point that departs from the above relation is due to gas/material releases from the radiation source. These releases lead to a repulsive force. Baking the lamp+foil is therefore essential for the experiment. Nevertheless, there should be no problem to repeat the experiment, let alone to verify that electromagnetic radiation on a lead sphere leads to an attractive force.

In a companion paper (Lundin and Lidgren, 2010), we describe how e/m-wave ponderomotive forcing may apply to the above experimental findings. The theory is based on an analogy with ponderomotive wave forcing of magnetized plasma (Lundin and Guglielmi, 2006), the Miller force leading to attraction below-, and repulsion above the plasma gyro-frequency. The theoretical force derived ($3.8 \cdot 10^{-9}$ N) is close to the force measured in vacuum: $2.3 \pm 0.4 \cdot 10^{-9}$ N (summation method), and $2.9 \pm 0.5 \cdot 10^{-9}$ N (integrated from curve).

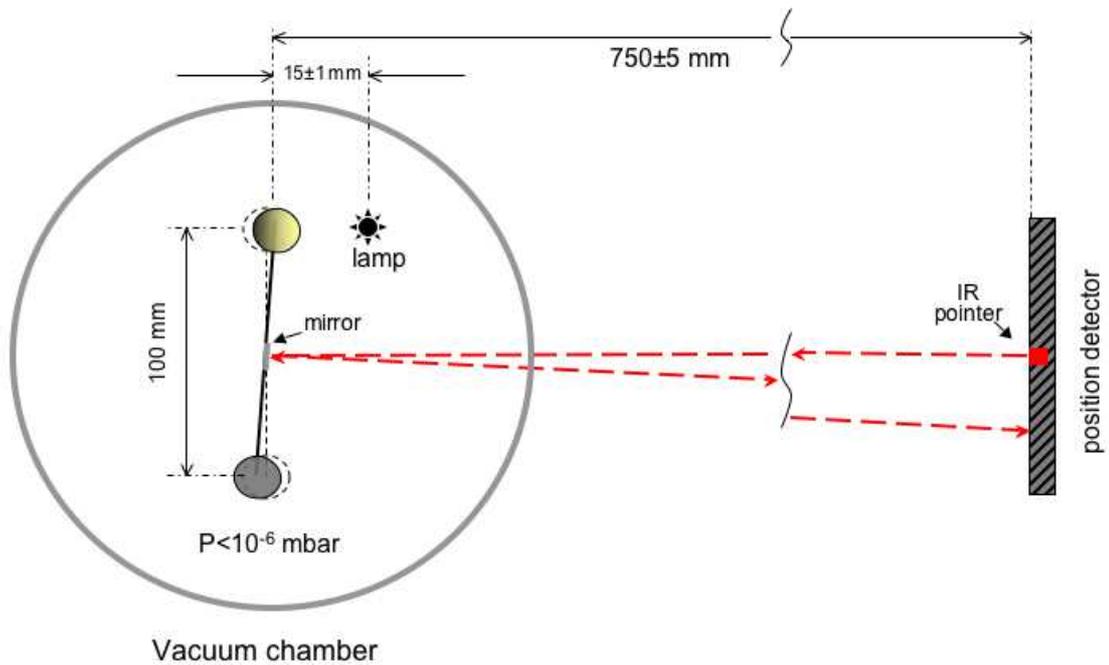

Figure 1. Diagrammatic representation of the Cavendish torsion pendulum placed in the vacuum chamber. The lamp covered by aluminium foil is placed inside the vacuum chamber in front of one of the lead spheres. The IR pointer and position detector is placed outside the vacuum chamber

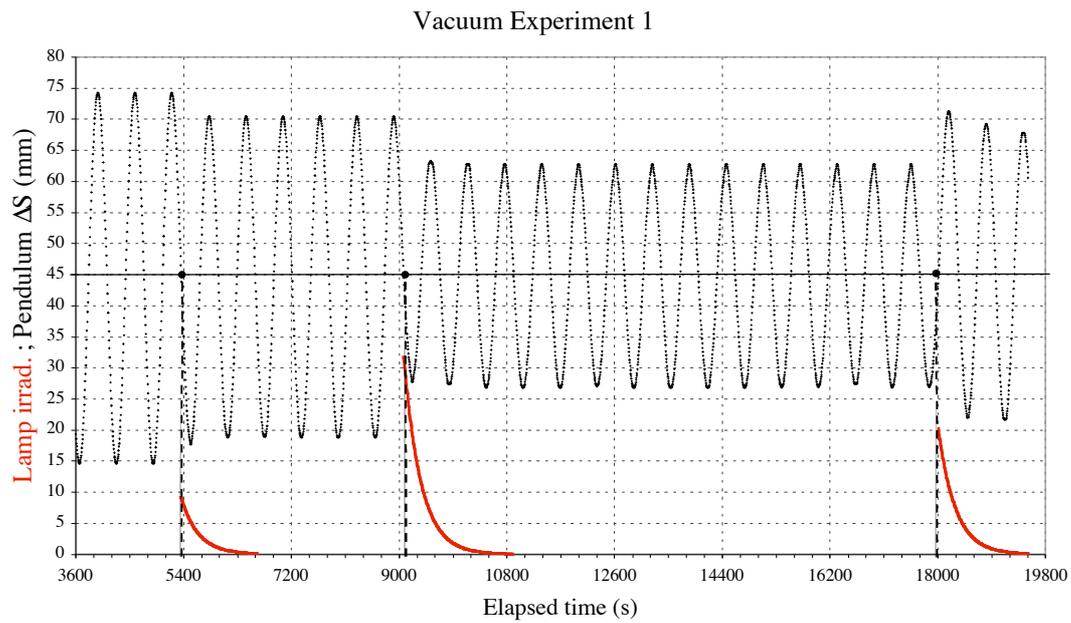

Figure 2. —The Cavendish V1 vacuum experiment. The experiment encompasses three lamp irradiation sequences (marked red), initiated at three different times (dots and dashed vertical lines). The sequence comprises irradiation by 9W for 10 s; by 26W during 10 s; and by 18W during 10 s. The figure illustrates clearly the effect of an attracting force during the irradiation periods.

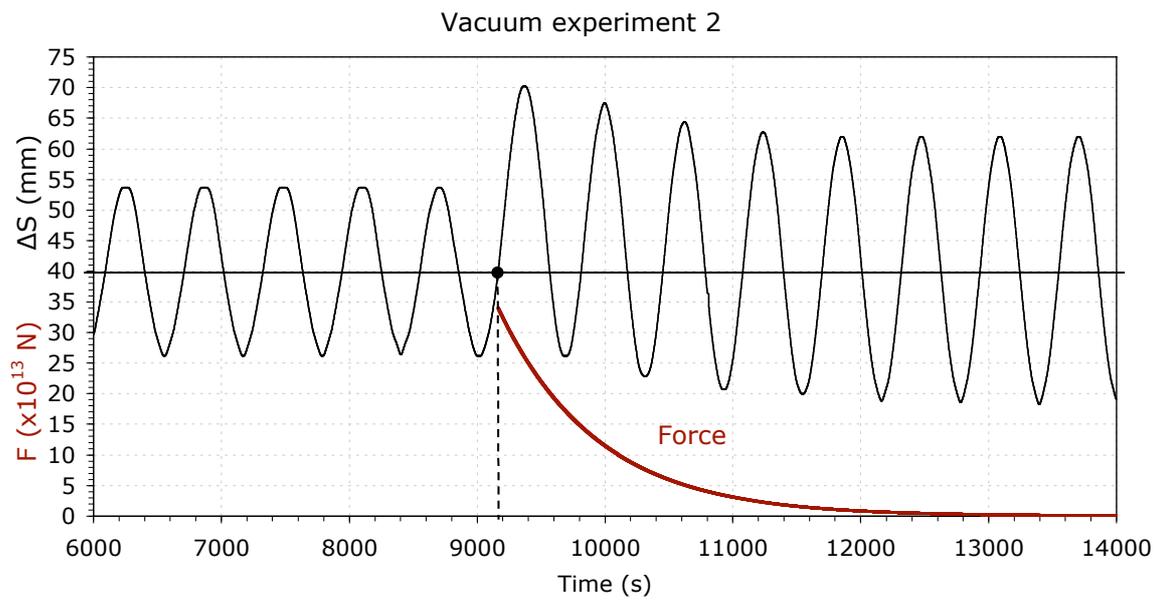

Figure 3. —The Cavendish V2 vacuum experiment. The force associated with the decaying radiation source, derived from the torsion pendulum displacement versus time, illustrated by the red curve.

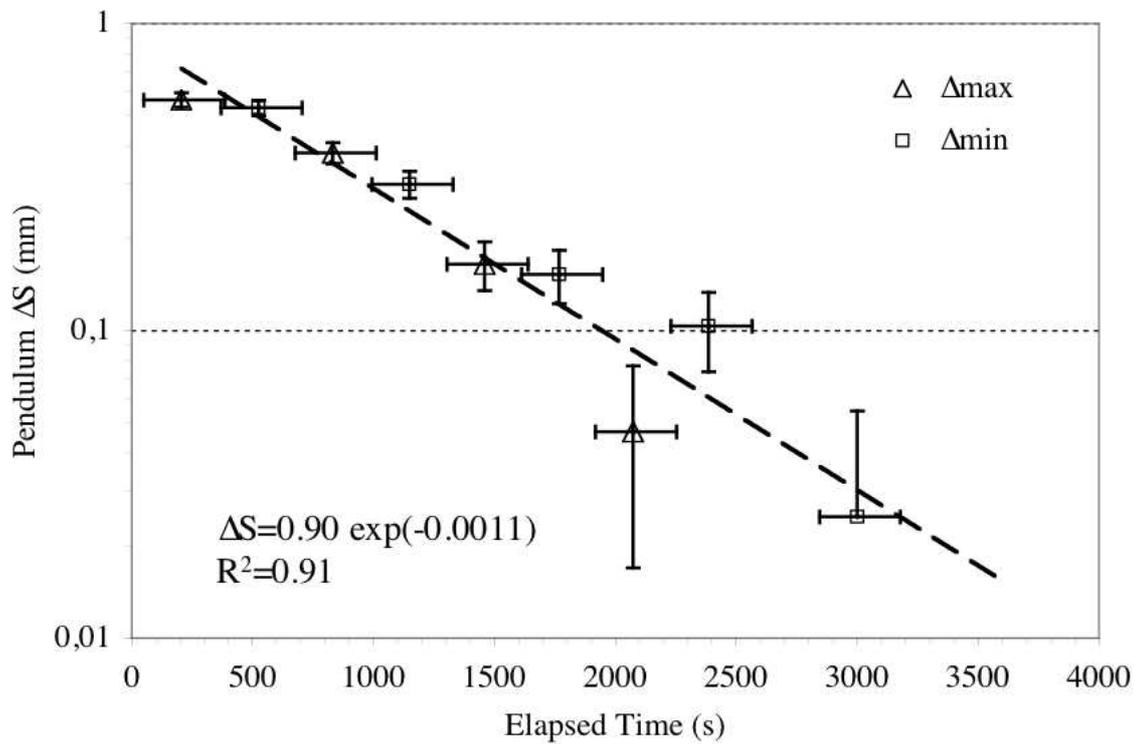

Figure 4  Pendulum displacement curve versus elapsed time for the V2 experiment. $\Delta$max and $\Delta$min mark the maximum and minimum pendulum offset ($\Delta S$) with respect to equilibrium. Dashed curve marks the least square fitted exponential function to the data points. Ordinate error bars mark the $\Delta S$ measurement error. The exponential displacement decay replicates the presumed exponential decay of the radiation source.

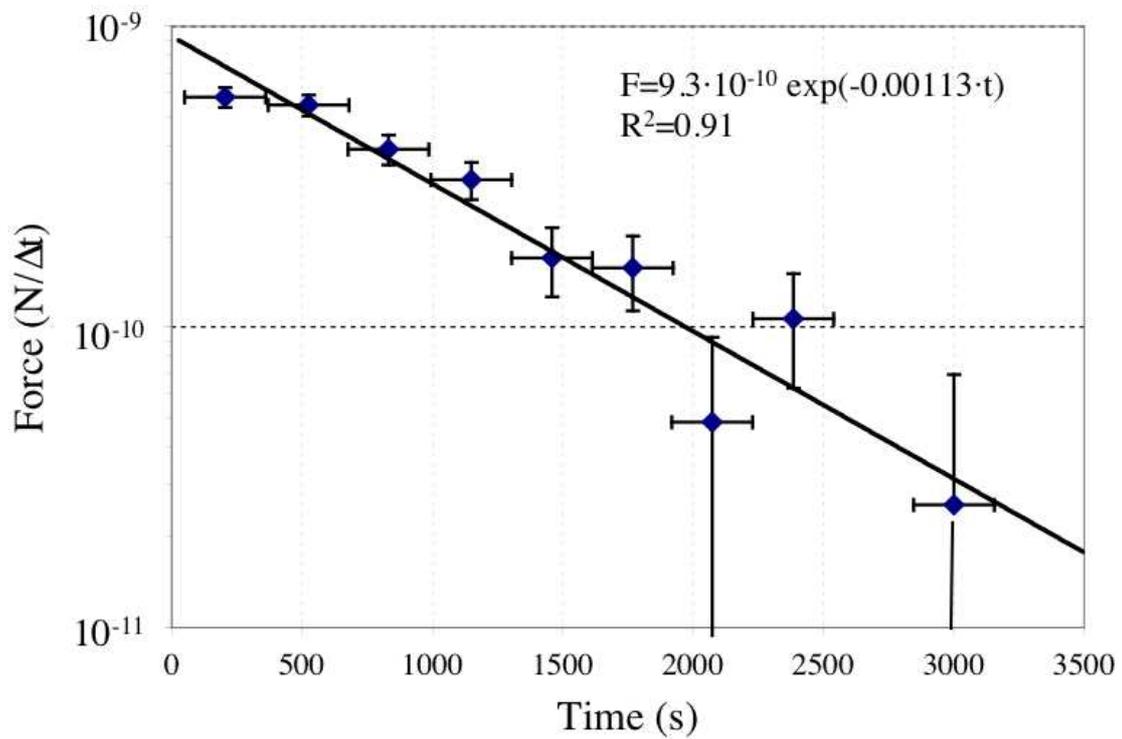

Figure 5. Force versus elapsed time derived from the average displacements of the pendulum amplitude during 1/2 cycle time intervals ($\Delta t$). Data points mark the accumulated force within a time interval $\Delta t$ (abscissa bar). Ordinate error bars correspond to the standard deviation offset from exponential function fit.

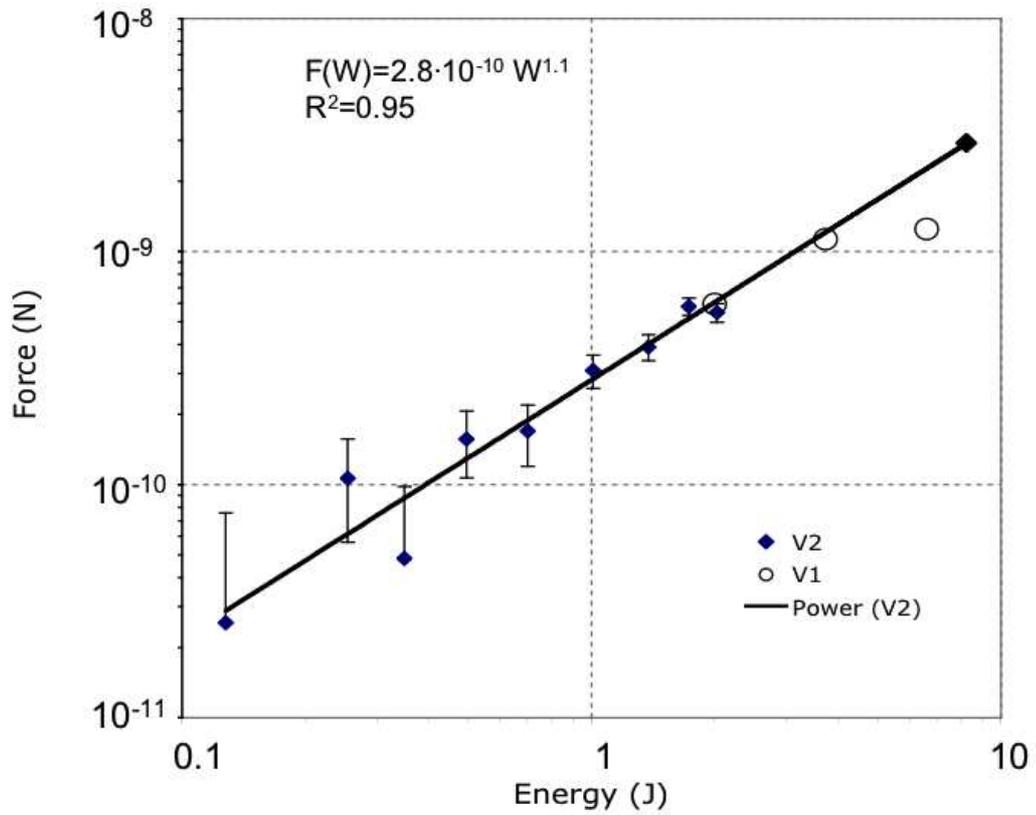

Figure 6 — Force versus total radiation energy affecting the target sphere in the V1 and V2 experiments. Black curve marks power law fit to the V2 data points. The difference between the V1 and V2 experiment is the "baking" preceding the V2 experiment. An opposing force, due to out-gassing from the lamp+aluminium foil, affects the V1 results (circles).